\documentclass[pra,twocolumn,floats,showpacs,preprintnumbers,tighten,epsfig,groupedaddress]{revtex4}

\usepackage{graphicx}
\usepackage{dcolumn}
\usepackage{bm}
\usepackage{amsmath}
\usepackage{amsfonts}
\usepackage{amssymb}

\usepackage{bm,color}

\begin{document}

\title{Developing an Action Concept Inventory}
\author{Lachlan P.~McGinness and C.~M.~Savage}
\email{craig.savage@anu.edu.au}
\affiliation{Physics Education Centre, 
Research School of Physics and Engineering,
Australian National University,
ACT 2601, Australia}

\date{\today}

\begin{abstract}
We report on progress towards the development of an Action Concept Inventory (ACI), a test that measures student understanding of action principles in introductory mechanics and optics. The ACI also covers key concepts of many-paths quantum mechanics, from which classical action physics arises. We used a multi-stage iterative development cycle for incorporating expert and student feedback into successive revisions of the ACI. The student feedback, including think-aloud interviews, enabled us to identify their misconceptions about action physics.
\end{abstract}

\pacs{01.40.G-, 01.40.Di, 01.40.Fk, 01.40.gf}
\maketitle
\preprint{Version: Draft \today }

\section{Introduction}
\label{Introduction}
Action physics has been included in first-year university physics courses at the Australian National University for a number of years, broadly following the model advocated by Edwin Taylor \cite{Taylor2003}. Specifically, we have taught Fermat's principle of stationary time for ray optics and Hamilton's principle of stationary action for mechanics. A first year text that introduces these topics is the Feynman Lectures on Physics \cite{Feynman Lectures}. Following Feynman's book ``QED'' \cite{QED}, Taylor \cite{Taylor2003,TaylorOersted}, and Ogborn and Taylor \cite{Ogborn}, we justified and motivated classical action physics from many-paths quantum mechanics.

There are few models for teaching action physics in a first-year university course, and hence taking a Physics Education Research (PER) approach to evaluation was important to determine the effectiveness of the innovation. Hence, we are developing an Action Concept Inventory (ACI), the current version of which is discussed in this paper and available in the Supplemental Material \cite{Supplemental Material}. As will be discussed in Sec.~\ref{Results}, this version does not meet the internal validity criterion for evaluating understanding of a single construct, such as ``action physics''.

Concept inventories are multiple-choice tests designed to efficiently evaluate students' conceptual understanding of a particular area of learning \cite{Inventory list}. They have been used in many areas of physics, most famously in Newtonian mechanics, where the Force Concept Inventory has had a major impact on teaching and learning \cite{Hestenes92, Hestenes95, Human,Hake}. 

There are  well developed methodologies for constructing and validating concept inventories \cite{Adams}. One of the first steps  is undertaking a literature survey of PER on the inventory's subject. However, as action physics is a relatively new area in PER, this literature is largely confined to content and its delivery \cite{Hanc2003,Moore,HancAJP2004a,HancAJP2004b,Neuenschwander}. There is little discussion of the student misconceptions that might inform the inventory development. Hence, a significant part of the work we report is the identification of misconceptions about action physics.

\begin{figure}[t]
\includegraphics[width=8cm]{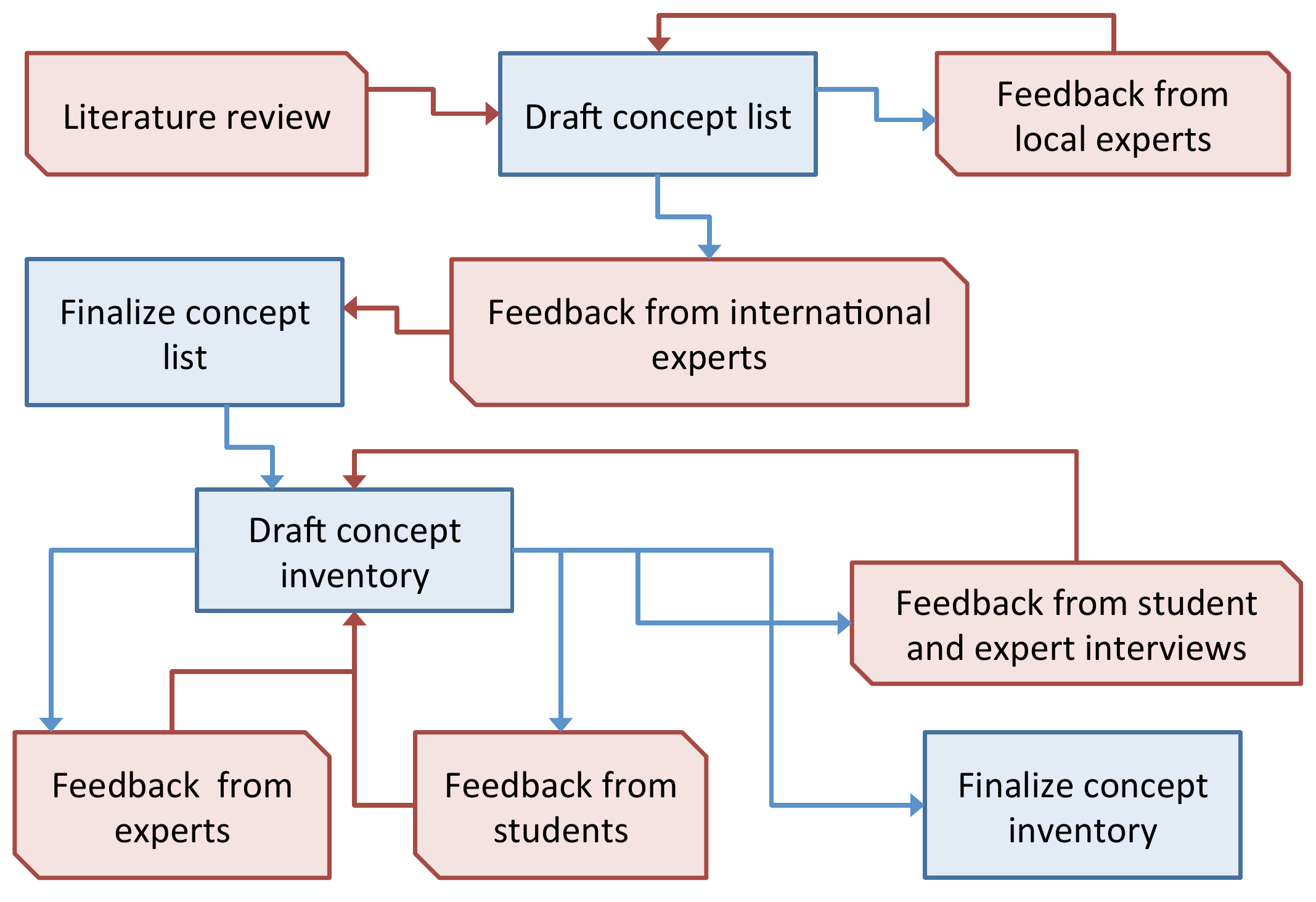}
\caption{The process for developing an Action Concept Inventory, showing feedback. Blue rectangular boxes represent creative processes while red truncated-corner boxes represent feedback and modification.}
\label{CI development}
\end{figure}
\begin{table*}
\caption{\label{concepts}The concepts tested by the ACI. In the questions column are the question numbers we classified as associated with each concept. The questions are available in the supplemental material \cite{Supplemental Material}.}
\begin{ruledtabular}
\begin{tabular}{p{6 cm}  p{9 cm} p{2 cm} }
  Concept & Description & Questions \\ \hline
Path & A path is a curve in space, with each of its points occurring at a specific time. & 10, 11 \\

\raggedright Principle of Stationary Action & The classical (observed) path is the one for which the action is stationary. & 7, 9, 16 \\

Stationary &A first order variation of the path makes no first order difference to the action. & 4, 12, 13 \\

Fermat's Principle & When travelling from point A to point B a light ray will take the path for which the time is stationary. & 6 \\

\raggedright Conservation of Energy & Energy is conserved. & 1, 2 \\

Principle of Stationary Potential Energy & A particle placed at rest at a point of zero slope in the potential energy curve will remain at rest. & 3, 5 \\

\raggedright Explore All Paths & In quantum physics a particle takes all possible paths when moving between two states. & 8, 19 \\

\raggedright Complex Amplitude & Each path has a complex amplitude determined by the action, with equal magnitudes, and different phases. & 15, 18 \\

Superposition Principle & Total amplitude for a transition from A to B is obtained by summing the amplitudes of each path from A to B.  & 14, 15, 17, 18 \\

Probability & The probability of a transition is calculated by squaring the modulus of the total amplitude. &  14, 17 \\

\end{tabular}
\end{ruledtabular}
\end{table*}
%

Including topics in first year courses that are usually taught in later years, such as action physics, is not to be undertaken lightly. It requires a level of student preparation and sophistication that will not be found in all classes. Our decision to teach action physics to first-year students, is based on the fact that our students are high achievers, according to the Australian university entrance score measure (ATAR), and that ANU Physics aims to educate to a very high standard \cite{PECMission}. In this paper we do not attempt to make the case for the broader adoption of such teaching, as we have done this elsewhere \cite{Action Course}. Rather we describe an evaluation tool to assist those who may wish to pioneer first year action physics instruction. 

In the next section we describe the ACI development process, including the core concepts identified in collaboration with experts.  Section \ref{Results} presents analyses of the results of administering the inventory. Section \ref{Misconceptions} discusses the misconceptions that we identified in the inventory development process. We conclude by touching on some questions regarding the teaching of action physics, and the role of the ACI.

\section{Developing The Action Concept Inventory}
\label{Developing The Action Concept Inventory}

To develop the ACI we used a multi-stage iterative cycle of the kind described by Adams and Wieman \cite{Adams}, and outlined in Fig.~\ref{CI development}. This approach is designed to be objective. It is based on advice from independent experts concerning the choice of concepts and questions, as well as on feedback from students. 

The first stage was to identify the key concepts of action physics by a literature review and by consultation with five Australian National University (ANU) academic staff who have taught action physics at various levels.
Then fifteen international experts were identified through their publications on action physics education. They reviewed the preliminary list of key concepts using an online survey; this resulted in additions, deletions and refinements. The ten concepts resulting from this process are given in Table \ref{concepts}. They were each supported by at least 70\% of the experts. The last four of the concepts concern many-paths quantum mechanics, the conceptual basis on which our instruction was built.

Examples of concepts eventually excluded from the list were: action, generalized coordinates, and Noether's theorem. Generalized coordinates and Noether's theorem were rejected as being non-fundamental extensions to the principle of stationary action. Action was excluded because it was judged to be more a definition than a concept. Its conceptual aspect is captured in part by the principle of stationary action concept.

In order to test understanding of these concepts, first-draft questions were developed for the inventory. We then conducted and analyzed thirty-one student think-aloud interviews using these questions. In think-aloud interviews students explain their reasoning while answering the open-ended questions, with the interviewer avoiding influencing the students' thinking. These students were a diverse group, nineteen of whom were later year undergraduates. Records of the interviews were analyzed for misconceptions, and hence for distractors, which informed our development of multiple-choice answers.

The resulting second-draft of the inventory, consisting of 18 questions with multiple-choice answers, was administered to the 2014 ANU ``Physics 2'' first-year university class before and after instruction. Physics 2 is the second course required for physics majors at the ANU. It constitutes a quarter of a full-time load for a semester. The 2014 class enrolment was 108, somewhat more than half of whom will complete three-year physics majors. Most students had completed ``Physics 1'' the previous semester, which reviewed the Newtonian mechanics and electromagnetic statics already studied at secondary school level. Physics 2 mathematics co-requisites include: complex numbers, partial differentiation and differential equations. The 2012 class was the subject of a similar study on relativity physics education \cite{Aslanides}. 

The action physics instruction was a three-week module of eight 50 minute lectures, a 90 minute computational laboratory, and two 50 minute tutorials. The lectures were active in style and included class-response questions. Students were provided with a set of notes, recommended reading \cite{Feynman Lectures,QED} and video viewing \cite{Cassiopeia}. Assessment was by two homeworks, a lab logbook and a question on the final exam. Student assessment work was part of the raw data for our research \cite {ethics}.

Twenty validation interviews were conducted after the post-instruction administration. As a result problems with certain questions were identified and corrected.  

Three of the validation interviews were conducted with pairs of students, who discussed their reasoning while being passively observed. This was a more comfortable environment for the students, and the discussion reduced the number of superficial errors in interpretation. It would be interesting to investigate how administering concept inventories to pairs of students might affect their accuracy as instruments for measuring class learning.

Next, four international experts reviewed the inventory, which led to further improvements. Finally, we added an additional question addressing the explore all paths concept (question 19 \cite{Supplemental Material}). The resulting inventory is presented in the supplemental material \cite{Supplemental Material}. 

\begin{table}
\caption{\label{ACI test stats}ACI post-test statistics for the 2014 and 2015 classes. Sample sizes were $N=76$ and $N=87$ respectively.}
\begin{ruledtabular}
\begin{tabular}{lccc}
  Statistic & 2014 & 2015 & Desired range \\ \hline
Mean item difficulty index & 0.35 & 0.43 & [0.3, 0.9] \\
Ferguson's delta & 0.93 & 0.95 & $\ge$ 0.9 \\
Mean discrimination & 0.24 & 0.38 & $\ge$ 0.3 \\
KR20 & 0.53 & 0.64 & $\ge$ 0.7
\end{tabular}
\end{ruledtabular}
\end{table}

\section{Results}
\label{Results}

Various forms of statistical analysis were applied to the action concept inventory response data. Some test statistics are given in Table \ref{ACI test stats}. The sample in each year is all students who completed both the pre and post-instruction tests. Students were given 30 minutes at the end of class to complete the ACI using an online form, with most students taking between 15 and 25 minutes. Students who were absent were encouraged to complete it out of class. 

The statistics are defined and explained by Ding and Beichner \cite{Ding}, from where the desired ranges are also taken. The mean item difficulty index is just the fraction of answers that were correct. Ferguson's delta is an indicator of the range of students' scores across the class. Discrimination is a measure of how well a question distinguishes between top and bottom quartile students, as determined by total test scores. In 2014 the ACI's mean discrimination was below the desired range, indicating that the difference in its difficulty for the top and bottom quartile students was less than recommended. This can happen because questions are too difficult or too confusing. After modifications to a number of questions the mean discrimination of the 2015 ACI moved into the desired range. 

Questions with particularly low discrimination, and difficulty indices - that is, particularly difficult questions - were four questions   testing three of the quantum mechanical concepts (Q14, Q15, Q17, Q18 \cite{Supplemental Material}), namely: complex amplitude, superposition principle, and probability. Evidence from the student interviews and homework confirmed that our instruction in the quantum mechanical concepts did not lead to students mastering them.

\begin{figure}[]
\includegraphics[width=8cm]{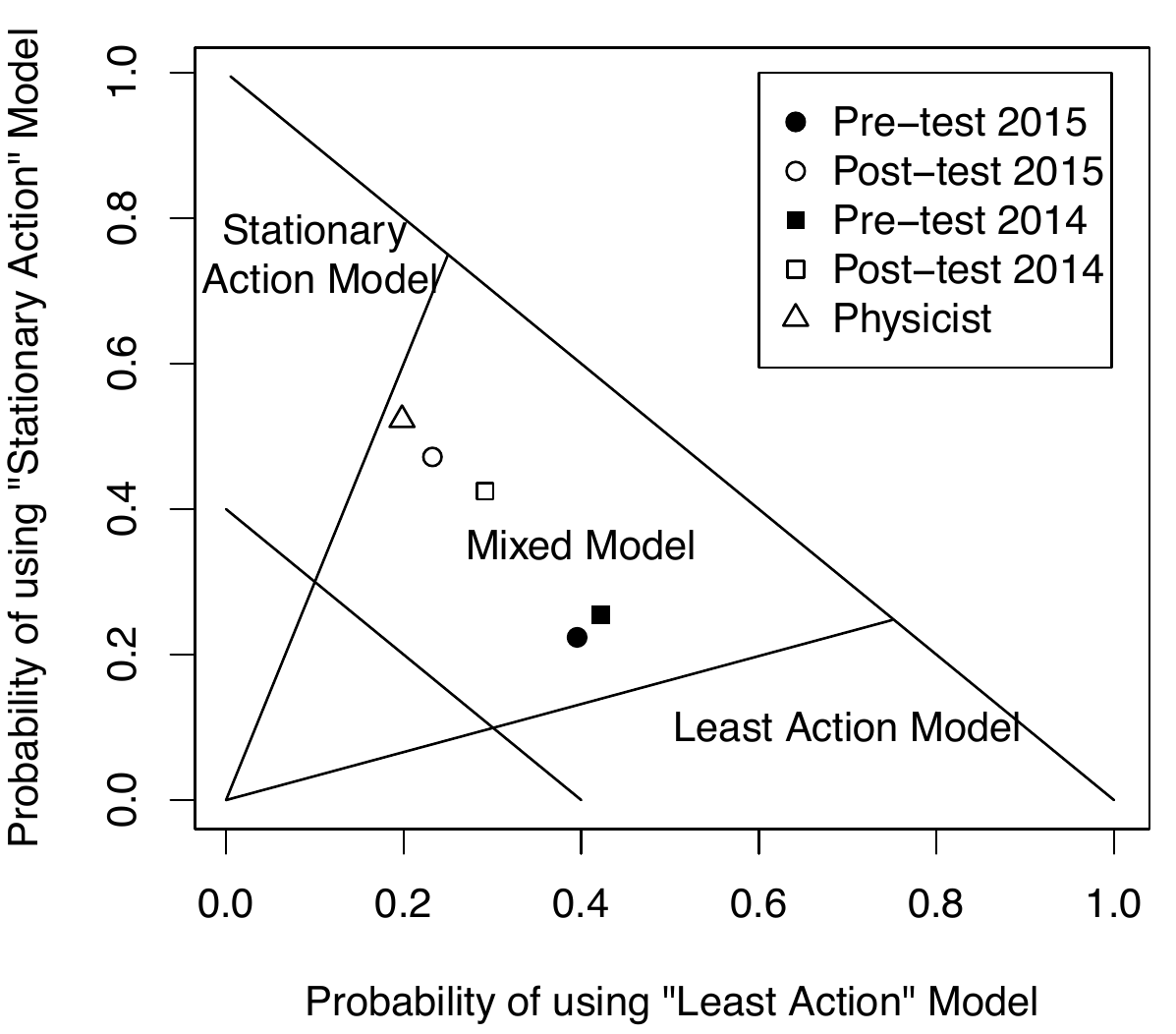}
\caption{Model plot of the probability of students using the correct principle of stationary action conceptual framework versus their probability of using the incorrect least action framework. The lines somewhat arbitrarily separate the plot into three regions corresponding to different conceptual frameworks \cite{Bao}:  they are where the sum of the two probabilities are 0.4 or 1, and where one probability is three times the other. All results are in the mixed model region, meaning that individuals were using either the stationary action or least action models, depending on the question. }
\label{model plot}
\end{figure}
%

The KR-20 test statistic is a measure of the consistency of the concepts measured by the inventory. It measures the consistency  of the different questions against each other. For inventories, such as the ACI, that measure a large number of concepts (10) relative to the number of questions (19) the KR-20 is also a measure of ``task variability'' \cite{Adams}. That it is below the desired range \cite{Ding, Maloney} indicates that the concepts being tested, Table \ref{concepts}, have a high degree of independence, and that there is no single idea, or construct, whose understanding is being tested by the ACI \cite{Lasry}. Hence it should not be regarded as a fully developed concept inventory, even though it is useful for other evaluation purposes, such as identification of misconceptions. Adding more questions, thus increasing the ratio of questions to concepts, may move the reliability statistic into the acceptable range.

For the 2014 class, the pre and post-instruction class average scores out of a possible 18 were $5.1 \pm 0.2$ and $6.3  \pm 0.3$, where the uncertainties are standard deviations of the means. The corresponding normalized gain, that is the actual post instruction gain as a fraction of the maximum possible gain, is $0.09 \pm 0.03$. The inventory thereby diagnosed that our teaching effectiveness for action concepts was low, with the quantum mechanical concepts being particularly problematic.

Hence, we increased our emphasis on quantum concepts in the 2015 Physics 2 class. Specifically, we increased the lecture and computational laboratory time spent on quantum mechanics, and emphasised the importance of studying the assigned introductory videos before the lectures \cite{Cassiopeia}. For the 2015 class the pre and post-instruction average scores out of a possible 19 were $5.2 \pm 0.2$ and $8.2 \pm 0.3$, for a normalized gain of $0.22 \pm 0.03$. 


We performed a model analysis of our 2014 and 2015 data. This is a method due to Bao and Redish \cite{Bao} for identifying students' conceptual frameworks from concept inventory data \cite{Smith, Rakkapao}. It is motivated by cognitive science evidence that, whereas an expert has only one framework, a student may have different frameworks whose activation is determined by the context. 

We applied it to frameworks for understanding the principle of stationary action concept, which was tested by three inventory questions (Q7, Q9, Q16) \cite{Supplemental Material}. In this case two models predominate and the data can be represented on a model plot, Fig.~\ref{model plot}  \cite{Bao}. The three labeled regions correspond to students dominantly using the correct ``stationary action'' model, the incorrect ``least action'' model, and the mixed model region in which either was used depending on the question. The least action model is based on the misconception that only minima are stationary cases. The model plot shows that both before and after instruction, individual students were in the mixed model region. However after instruction the correct model was more likely to be used.

Applying model analysis to the explore all paths concept showed that the correct ``explore all paths'' model and the incorrect ``single path'' model dominated students' thinking. It also showed that our instruction was ineffective in shifting students towards the correct model, indicating the need for teaching improvements.

As an adjunct to our central project of measuring student learning, we investigated whether administering the ACI to physicists with advanced training might contribute to its development. Hence, we administered it to 61 ANU physicists: 22 were academics, 23 Ph.D. students, and the remaining 16 were third or fourth year undergraduates, or masters students. For most questions this group did significantly better than the first-year students. Two of the questions they did worse on had already been identified as needing revision as a result of the post-test student interviews, suggesting that expert groups can help identify question validity. This has the potential to accelerate the development process, given that classes usually run on a fixed schedule, while physicists are often available over a wider range of times.

Interestingly, the physicists did much worse than post-instruction Physics 2 students on one of the questions concerning the path concept, even though the question was valid (question 10 \cite{Supplemental Material}). This indicated that the physicists were not familiar with the concept that the boundaries of paths are events, specified by a time and a place. The distinction between the action physics concept of a path with two boundaries, and the conventional concept of a trajectory with initial conditions is a fundamental one. In contrast, Physics 2 students had repeatedly calculated actions along paths and hence many had mastered the concept.

\section{Misconceptions}
\label{Misconceptions}

In this section we discuss students' misconceptions about action physics that were revealed by our research. We also suggest how they might be addressed by instruction. Misconceptions were investigated using student interviews and conceptually oriented assessment tasks.  As the research progressed, our evolving understanding of misconceptions was used to iteratively improve the distractor multiple choice answers in the inventory, as previously described. 

Students completing the inventory were also asked to rate their confidence in their answer to each question \cite{Aslanides}. The purpose of this was to help identify misconceptions: an incorrect answer together with high confidence suggests that answer corresponds to a misconception. The student interviews revealed that some students who were clearly guessing answers would nevertheless rate their confidence in their answer as high. Hence we only used the confidence data to identify potential misconceptions after transforming it to a z-score for each student, that is subtracting the student's mean confidence over all questions and dividing by the standard deviation.

One misconception about the principle of stationary action concept is the idea that only minimum action paths are physical. This is an example of where the associated distractor was assigned high confidence by many students who chose it. Another misconception is that only one path can have stationary action \cite{more than one path,Gray}. 

Students commonly misconceived paths in mechanics as spatial paths, rather than as space-time worldlines. In the case of optics, the paths are spatial, but not in mechanics \cite{Maupertuis}. It is a challenge for the instructor to clarify not only the unity of optics and mechanics within action physics, but also their specific differences.

The concept of different orders of variation challenged many students. One misconception was that if the first order variation is zero, at a stationary point, then there is no variation at all, e.g.~that the second order variation is also zero. A specific example of this misconception is the belief that a path near a stationary path has the same action as the stationary path. Another misconception was that (non-zero) first order variations might be smaller than second order variations. This is perhaps the result of focussing on the magnitudes of the first and second derivatives in the relevant Taylor expansion, rather than on their respective products with the infinitesimal expansion powers $\epsilon$ and $\epsilon^2$.

The explore all paths concept was particularly difficult for some students to learn. This was expected as it is a deeply quantum mechanical idea, inconsistent with everyday experience. In applying classical conceptions some students believed that there is only one path that particles can take. Others believed that although different paths were possible, only one was physically relevant. Instruction should acknowledge the conflict with intuition while explaining that microscopic systems, of which we have no direct experience, have simply been found to be different.

Some students struggled to relate the concept of a path's complex amplitude to constructive and destructive interference in the amplitude sum. Even though almost all students had studied complex numbers in a mathematics course, there was sometimes a misconception that more paths always means a higher magnitude for the amplitude sum. We also found that some students assumed that the magnitude of the amplitude of the stationary path is greater than that of other paths, even though the magnitude of the amplitude of all paths is the same. This misconception focusses on individual paths rather than on the collective contribution of groups of paths.

The most common misconception regarding probability that we found was famously published in Max Born's 1926 paper \cite{Born}. This misconception is that the probability is the magnitude of the complex amplitude, rather than its magnitude squared. Born inserted a famous footnote in proof, correcting this misconception. An English translation is: ``More careful consideration shows that the probability is proportional to the square of the quantity $\Phi_{n,m}$''. It is not too much of a stretch to suggest that this correction won him the Nobel Prize in 1954 for his interpretation of the wavefunction.

\section{Conclusions}
\label{Conclusions}

Two important questions have not been addressed in this paper, as we have done so elsewhere \cite{Action Course}: \textit{should} and \textit{can} action physics be taught at first year university level? The \textit{should} case has been argued by Taylor \cite{Taylor2003,TaylorOersted} based in part on the unifications it enables. Whether or not we \textit{can} teach action physics effectively will depend on the students' preparation and on the effectiveness of innovative techniques, such as interactive visualizations \cite{TaylorCIP}, for addressing gaps in preparation. The ACI is a tool for investigating the \textit{can} question by measuring instructional success. 

The systematic and iterative concept inventory development process we have described has enabled us to improve the teaching of action physics at first-year university level. In particular, it has enabled us to identify the quantum mechanical foundation as a weak point in our instruction, and to redesign the Physics 2 course to improve it.



\begin{thebibliography}{99}
\bibitem{more than one path} An example of more than one path being stationary is provided by the one-dimensional harmonic oscillator. For initial and final events at the potential minimum, with the final event a half period after the initial event, there are stationary paths corresponding to motion to the left and to the right.
\end{thebibliography}
\end{document}